# Ferromagnetism in defect-ridden oxides and related materials

J.M.D. Coey, P. Stamenov, R. D. Gunning, M. Venkatesan, K. Paul
School of Physics and CRANN,
Trinity College, Dublin 2, Ireland

**Abstract.** The existence of high-temperature ferromagnetism in thin films and nanoparticles of oxides containing small quantities of magnetic dopants remains controversial. Some regard these materials as dilute magnetic semiconductors, while others think they are ferromagnetic only because the magnetic dopants form secondary ferromagnetic impurity phases such as cobalt metal or magnetite. There are also reports in $d^0$ systems and other defective oxides that contain no magnetic ions. Here, we investigate $TiO_2$ (rutile) containing 1 - 5% of iron cations and find that the room-temperature ferromagnetism of films prepared by pulsed-laser deposition is not due to magnetic ordering of the iron. The films are neither dilute magnetic semiconductors nor hosts to an iron-based ferromagnetic impurity phase. A new model is developed for defect-related ferromagnetism which involves a spin-split defect band populated by charge transfer from a proximate charge reservoir –in the present case a mixture $Fe^{2+}$ and $Fe^{3+}$ ions in the oxide lattice. The phase diagram for the model shows how inhomogeneous Stoner ferromagnetism depends on the total number of electrons $N_{tot}$, the Stoner exchange integral $I$ and the defect bandwidth W; the band occupancy is governed by the *d-d* Coulomb interaction *U*. There are regions of ferromagnetic metal, half-metal and insulator as well as nonmagnetic metal and insulator. A characteristic feature of the high-temperature Stoner magnetism is an an anhysteretic magnetization curve which is practically temperature independent below room temperature. This is related to a wandering ferromagnetic axis which is determined by local dipole fields. The magnetization is limited by the defect concentration, not by the 3*d* doping. Only 1-2 % of the volume of the films is magnetically ordered.

# 1. Introduction

The magnetism of thin films of transparent oxides doped with a few percent of $3d$ ions [1,2] has been a subject of intense controversy ever since the first report of high temperature ferromagnetism in thin films of cobalt-doped anatase by Matsumoto in 2001 [3]. There are similar reports of unexpected weak ferromagneticm in other '$d^0$ systems' which contain no $3d$ ions [4-6]. Almost anhysteretic magnetization curves have been reported for many other films such as ZnO:Co [7], $SnO_2$:Fe [8], $Cu_2O$:Co [9] and $In_2O_3$:Cr [10]. The $TiO_2$:Fe system has been shown to be magnetic in several previous investigations [11]–[18], although the magnetism of highly-perfect films was due to a secondary $Fe_3O_4$ phase [13]; the ferromagnetism of Cr-doped $TiO_2$ films is found to disappear as the crystalline quality improves [2,32].

Long-range magnetic order in dilute magnetic oxides is puzzling; the concentration of magnetic dopant cations lies far below the percolation threshold, and normal superexchange and double exchange interactions couple nearest-neighbour cations only. Some of the examples are insulating [19,20], so exchange mediated by mobile carriers cannot be a general explanation. Furthermore the exchange energy density at a few percent doping is far too small to produce ferromagnetism at room-temperature [21], even if the exchange were as strong as it is in cobalt, the ferromagnet with the highest known Curie temperature. We really should not be seeing *uniform* ferromagnetism at temperatures of many hundreds of degrees in these materials. Another problem is that the observed magnetization sometimes exceeds the maximum value that can be associated with ferromagnetically-aligned dopant ions [15,22,23]. The Ferromagnetic $d^0$ systems [4]-[6], [24]-[29], where there are *no* magnetic dopants, are the most extreme examples.

Bulk samples [30,31] and thin films [32] of the dilute magnetic oxide solid solutions behave as expected when they are well-crystallized and stoichiometric - they are paramagnetic down to the helium temperature range. Hence, there is an supposition that the ferromagnetic behaviour of thin film or nanoparticle samples must be due to ferromagnetic impurity phases present as inclusions [13,33] or as uncontrolled contamination, introduced, for example, by handling samples with steel tweezers [34,35]. The magnetic moment of a typical thin film sample, $10^{-8}$ A m$^2$, corresponds to that of a speck of iron or magnetite a few tens of microns in size. Clearly, care is needed to avoid the experimental pitfalls associated with measurements of very small moments [35,36], whether with a SQUID or an alternating-gradient force magnetometer. It is also essential to run proper control samples.

We have investigated $TiO_2$ films which were produced in oxidizing or reducing conditions and doped with different levels of iron. Here we focus on the oxidized films which turn out to have a significant magnetic moment, although the films contain no iron-based ferromagnetic impurity phase. The undoped films are not magnetic. Conversion-electron Mössbauer spectroscopy and magnetic susceptibility measurements show that iron is not directly involved in the ferromagnetic ordering. Following this experimental demonstration, we then develop a new general model based on our idea of charge transfer ferromagnetism [37]. Our previous model for dilute magnetic oxides [21] was a Heisenberg model where ordered moments residing on the 3$d$ dopant ions interact via magnetic polarons. The present model which we develop here is a Stoner model, where the 3*d* ions act as a charge reservoir, but do not necessarily contribute to the magnetic order. A defect based impurity band is *spontaneously* spin split. We show how the model works, and we calculate its properties at $T = 0$ K. The phase diagram is rich with high-temperature ferromagnetism appearing in the half-metallic regions.

**2. Experiment**

A series of films was prepared by pulsed-laser deposition from sintered targets made from 99.999 % pure $TiO_2$ and 95% isotopically-enriched $^{57}Fe_2O_3$. The targets had an Fe/(Fe+Ti) atomic ratio in the range 0 – 5 at %. Substrates were 5 x 5 x 0.5 $mm^3$ squares of R-cut ($1\bar{1}02$) sapphire, polished on both sides. They were maintained at 750°C during deposition, in an ambient oxygen pressure of $10^2$ or $10^{-3}$ Pa, using a KrF excimer laser operating at 248 nm and 10 Hz. Laser fluence on the target was 2 J $cm^{-2}$. The resulting films were 70-140 nm thick, with an rms roughness of 2.5 nm. Magnetization measurements were made in a 5 T SQUID magnetometer, by mounting the samples in drinking straws with no other sample holder. The $^{57}$Fe Mössbauer spectra were collected at room temperature in transmission geometry with a $^{57}$Co (Rh) source. Optical spectra were recorded at room temperature using a Perkin Elmer dual-beam spectrophotometer.

**3. Results and discussion**

Typical X-ray data are shown in Fig 1. The films have the rutile structure with a <101> texture. The only perceptible secondary phase in any of the oxidized films deposited at $10^2$ Pa was antiferromagnetic α-$Fe_2O_3$, in films with 5 % iron. Compositional analysis by EDAX reveal compositions which are normally within 10 % of the nominal one. The optical bandgap for the oxidized films was close to the 3.1 eV expected for rutile $TiO_2$ [38], whereas the bandgap of films produced in reducing conditions was 3.9 – 4.2 eV, due to the formation of a Magnéli phase. For the latter films, there is some indication of impurity absorption below the band edge (400-500 nm) for both undoped and iron-doped films.

All undoped samples (also Zn-doped samples) - eleven of them altogether - exhibited a straightforward linear negative susceptibility (Fig 2a), due to the diamagnetism of the substrate, $\chi$ = -4.8 x $10^{-9}$ $m^3$ $kg^{-1}$. We found no trace of the ferromagnetic component that has been reported for undoped anatase films deposited on $LaAlO_3$ [26, 27].

Two thirds of the films doped with $^{57}$Fe or natural Fe showed a ferromagnetic signal superposed on the diamagnetic background. Data shown in Fig. 2b are corrected for the high-field diamagnetism of the substrate and any paramagnetism of the sample. Saturation moments are in the range 2 –20 ×$10^{-8}$ $Am^2$ or 90-900 $\mu_B$ $nm^{-2}$ of substrate area. It can be seen several of them have moments in excess of 2.2 $\mu_B$ per iron, the value for iron metal; for example a moment as high as 6.9 $\mu_B$ per iron atom was found in the as-prepared oxidized 1 % film made with $^{57}$Fe.

In order to determine the magnetic state of the iron dopant, Mössbauer spectra were collected using a conversion electron detector. The mean escape depth of the 7.3 keV conversion electrons in $TiO_2$ is calculated to be about 140 nm [39], so they effectively probe the entire depth of the films. The whole of the 5 × 5 $mm^2$ is irradiated with γ-rays. The data with least-squares fits to the hyperfine parameters are shown in Fig 3, and the fit parameters are listed in Table 1. It is evident from these data that all the iron is *paramagnetic* in the 1 % or 3 % oxidized films and most of it is in the $Fe^{2+}$ or $Fe^{3+}$ state, respectively. The isomer shift and quadrupole splitting and are in the range for iron ions in octahedral oxygen coordination, and correspond fairly well to those reported for Fe-doped $TiO_2$ prepared by ball milling [40,41]. The spectrum for ferromagnetic iron metal, as shown in Fig. 3d, is quite distinctive, and unlike those of hematite or magnetite. A second phase (hematite) appears at 5% Fe, but not at 1% or 3%. We find no evidence for magnetite in any of the samples, and even if it were present, it can explain at most a moment of 1.33 Bohr magnetons per iron. We measure up to 6.9 $\mu_B$. Even allowing for differences in recoiless fraction and conversion efficiency, we are confident that at most 10 % of the iron in the 1% and 3% oxidized films could be present as metal. This means a moment of 0.2 $\mu_B$ per iron at most, which is very far from the measured values.

Further evidence that the iron remains magnetically disordered is provided by the high-field susceptibility, Fig 2d, which exhibits Curie-law behaviour consistent with the iron content of the film. Similar conclusions on the state of iron in $TiO_2$ films produced in high or low oxygen pressure were recently reached by Rykov et al [17]. Evidence is accumulating that the 3$d$ dopant in other ferromagnetic films, such as Co-doped ZnO, is not magnetically ordered either [42,43].

Nevertheless, the films of Fig. 3 exhibit ferromagnetic magnetization curves. Although the reduced magnetization curves measured between 4 and 300 K [Fig.2e] can be very roughly fitted to a Langevin function

$$M/M_0 = \coth y - 1/y \tag{1}$$

with $y = \alpha\mu_0 H$ where $\alpha = 2.8$ is a constant, the specimens are definitely not superparamagnetic. For a superparamagnet $\alpha$ must vary as $1/T$, but we find it is actually independent of temperature.

We now address the origin of the magnetism. Contamination is ruled out because undoped films prepared in the same chamber and handled in the same way as the doped films exhibit no moment. (Our detection limit for magnetic moment is better than $10^{-9}$ A m$^2$). Of the 16 iron doped films prepared in oxidizing or reducing conditions which showed ferromagnetism, for only one, a reduced film with 3% iron that differed from all the others in that it exhibited temperature-dependent hysteresis, could the moment be quantitatively explained by an α-Fe impurity. Four of the iron-doped films, all with x≤1%, were nonmagnetic.

Since the magnetic moments of the dopant iron ions are not magnetically ordered, they cannot be the direct source of the ferromagnetism; we need another explanation. The clue, highlighted by the 1% oxidised film, is the mixed valence state of the iron. The example of ilmenite, FeTiO$_3$, suggests that no reduction of Ti$^{4+}$ is to be expected in an oxide in the presence of Fe$^{2+}$. The effect of replacing Ti$^{4+}$ by Fe$^{2+}$ in the rutile lattice is the creation of electronic or structural defects. For example, the substitution

$$\text{Ti}^{4+} + 2\text{O}^{2-} \rightarrow \text{Fe}^{2+} + 2\text{O}^{-} \qquad (2)$$

creates two ligand holes in the *p*-shells of oxygen anions close to the iron cation. These holes will be strongly correlated, but delocalized over the first or more distant coordination spheres. Their spins will tend to couple ferromagnetically. Any orbital moments on the O$^-$ anions should couple parallel to the spin, according to Hund's third rule, but the spin-orbit coupling for oxygen is very weak. The orbital moment should be quenched by the crystal field, but in any case, the greatest moment achievable for the $^2P_{3/2}$ state with g = 4/3 is 2 μ$_B$. The fact that we observe much larger moments in some samples suggests a role for the oxygen vacancies that have been reported in Fe-doped

TiO$_2$ [29,44] or other defects in quantities that significantly exceed the number of iron ions. A suggestion by Gallego et al [45] that magnetism in oxide films is due to oxygen holes associated with the surface atoms leads to a moment of 54 $\mu_B$ nm$^{-2}$ in TiO$_2$, assuming 1 $\mu_B$ per oxygen and taking both surface and interface oxygen into account. This is insufficient to explain the moments observed in the 1% samples. The defects would have to extend beyond a single surface layer. Another suggestion has been a giant paramagnetic orbital moment associated with surface currents [46,47], although such screening currents are usually diamagnetic.

**4. An explanation: Charge transfer ferromagnetism**

In formulating a new model for the magnetism of the doped oxides and other materials which unexpectedly turn out to be magnetic, we take a cue from the mixed valence of the iron dopant in the TiO$_2$ films. Ions coexisting in two different 3$d$ configurations constitute a *charge reservoir*, from which electrons may be easily transferred to or from states associated with defects in the lattice. The mechanism for magnetic order, which we call *charge transfer ferromagnetism* has three main components;

i) a defect-based band with a high density of states in the vicinity of the Fermi level

ii) the charge reservoir to or from which electrons can be easily transferred and

iii) an effective exchange integral *I* associated with the defect states.

The defect band, is closely analogous to the impurity band in semiconductors. Provided the effective Stoner integral *I* is large enough, the energy gain from exchange splitting the defect band can compensate not only the kinetic energy cost of splitting of the band, as in the standard Stoner model, but also the energy cost of transferring the electrons from the charge reservoir [37]. A new sort of ferromagnetism results. Unlike a classical Stoner ferromagnet, where the magnetization is uniform and simply determined by the density of states at the Fermi level $N(E_F)$ in the unsplit band, provided the Stoner criterion

$$I N(E_F) > 1 \tag{3}$$

is satisfied, a charge transfer ferromagnet is not uniformly magnetized. Only the regions containing the defects become ferromagnetic. Furthermore, the relevant density of states is not necessarily that *at* the Fermi level in the unsplit defect band. It would be a fluke for the Fermi level to fall near the maximum of the density of states. A high density of states above or just below the Fermi level will suffice, in conjunction with the charge reservoir.

Some possible distributions of defects are illustrated in Fig. 4. The defects might be distributed uniformly throughout the material (Fig.4a), or they may aggregate for example via spinodal decomposition (Fig.4b). They could be associated with grain boundaries (Fig.4d), or with the surface or interface of a thin film (Fig.4c), or with the surface of a nanoparticle. In this way, only a small fraction the total volume of the sample need be involved. Recent compilations of data on undoped and Mn-doped ZnO by Straumal *et al* show that the magnetization correlates with volume occupied by grain boundaries [48].

The physical mechanism is illustrated in Fig. 5. Electrons are transferred from the charge reservoir to the defect band, or vice versa, which has the effect of increasing $N(E_F)$ in the defect band to the point where the Stoner criterion (3) is satisfied. The shaded regions in Fig. 4 then become ferromagnetically ordered.

In order to explore the model of charge transfer ferromagnetism quantitatively, we consider the expression for the total electronic energy, based on three parameters $I/W$, $U/W$, and $N_{tot}/W$. $I$ is the Stoner integral in the defect band, $W$ is the defect bandwidth, $N_{tot}$ is the total number of electrons (including the band and the reservoir) and $U$ characterizes the charge reservoir. Here $U$ is taken to be the on-site Coulomb energy for transforming a dopant cation from $M^{n+}$ to $M^{(n-1)+}$ (i.e. converting $Fe^{3+}$ to $Fe^{2+}$). For computational convenience, and to illustrate the richness of the model, we choose a two-peaked defect density of states $N(E)$ for one spin state ,

$$N(E) = (15/4)[8(E-W)^2/W^3][1 - 4[(E-W)^2/W^2] \text{ for } 0 \leq E \leq 2W \quad (4)$$

Any one or many-humped density of states with at least one pronounced maximum will give similar results. Another example would be $N(E) = (2/W) \sin^2 (2\pi E/W)$ for $0 \leq E \leq W$.

There are three terms in the expression for the total energy, the band term, the exchange term and the charge transfer energy:

$$E_{\text{tot}} = \int_0^{E_0+\varepsilon} E' N(E') dE' + \int_0^{E_0-\varepsilon} E' N(E') dE' - I(n_\uparrow - n_\downarrow)^2 + U[(n_\uparrow + n_\downarrow) - n_0]^2 \quad (5)$$

where $2\varepsilon$ is the splitting of the defect band, $n_{\uparrow,\downarrow} = \int_0^{E_0 \pm \varepsilon} N(E') dE'$ and $n_0$ is a constant that reflects the 3d occupancy. $0 \leq n_0 \leq 2$. The problem is treated within the canonical ensemble with the Landau potential equal to $\Omega = E_{\text{tot}} - TS - \mu N_{\text{tot}}$. Here S is the entropy and $\mu$ is the chemical potential of the system (common to both the defect band and the charge reservoir). As a simplification the $\Omega$ potential is minimized in the limit $T = 0$ K, and the resulting system of two non-linear integral equations is solved numerically for $\varepsilon$ and $\mu$. The resulting solutions may not be unique, because the numerical solutions may depend on the initial guesses for $\varepsilon$ and $\mu$. However, the two-dimensional initial parameter space is sampled with a large number of points, and the solution (the pair ε, μ with the smallest free energy $E_{tot}$ ) is selected from among all resulting solutions. The resulting values of ε and μ are then used to evaluate various system parameters for particular choices of $U$, $N_{tot}$, $I$, $n_0$. The actual bandwidth W is not important as all the system parameters are normalized by it.

Fig. 6 is an example of a phase diagram, calculated for fixed values of $I/W = 1$ and $n_0/W = 0.4$, which illustrates the region of stability where spontaneous charge-transfer ferromagnetism occurs. It is favoured for intermediate values of $U/W$ and $N_{tot}/W$ and large values of the Stoner integral $I$. A number of different phases, both metallic and semiconducting can be identified. The Stoner integral $I$ fluctuates across the periodic table, but for light elements such as oxygen it is about 1 eV [49]. A smaller value may be expected for the free-electron gas. Ferromagnetism of an oxygen-based defect band is expected particularly in the coloured regions of Fig. 6. For example, $U$ should be less than 3 eV, the band-gap in $TiO_2$, so if $U = 0.5$ eV, $W$ must be ≤ 1 eV. This is a plausible value.

The double humped structure in the defect density of states gives rise to the rich phase diagram shown in the figure. It should be noted that a sequence of alternating magnetic and nonmagnetic phases is predicted with increasing electron occupation of the defect band. There is some evidence for this in the ZnO:Co system [50].

It may be noted that sequences of phases such as FI/NM/FM are predicted on Fig.6 with increasing electron concentration. This sequence has been observed in Co-doped ZnO [48]. Localized states due to potential fluctuations (Anderson localization) may be expected when there are few electrons in the impurity band. The charge transfer ferromagnetism model is not limited to dilute magnetic oxides. It can apply to $d^0$ ferromagnets as well, provided there is the requisite charge reservoir. Possible examples are the gold nanoparticles [4], germanium nanocaps [5], graphite [6], mentioned in the introduction, as well as oxide films such as $HfO_2$ containing cations with a high charge state [24], defective hexaborides [25], oxygen-irradiated $TiO_2$ crystals [29] etc. A characteristic of the ferromagnetism in all these systems is that it is a high-temperature phenomenon, which is practically anhysteretic. Both these features flow naturally from the model.

Were it not for spin wave excitations, Stoner ferromagnetism would lead to a Curie temperature of order the band splitting, which can be a fraction of an electron volt, or thousands of kelvins. The magnetism of a spin-split impurity band, without the charge reservoir, has recently been investigated by Edwards and Katsnelson [51] in the context of $d^0$ ferromagnetism in $CaB_6$. They find that the spin wave stiffness $D$ for a half-metallic impurity band can be of order $(1/12)R^2 W_v$ where R is the nearest neighbour distance and $W_v$ is the valence bandwidth. In $TiO_2$ the width of the 2$p$ band is 10 eV and R ~ 2 nm, D is therefore 3.3 eV nm$^2$, which is much greater than that of cobalt (5 meV nm$^2$). The spin wave excitations are therefore suppressed. High temperature ferromagnetism is therefore expected in the half metallic region of the phase diagram of Fig. 6. An experimental problem with measuring high Curie temperatures in these systems is that the defects are not in thermodynamic equilibrium, and their concentration evolves irreversibly on heating.

The screening of the on-site Coulomb interaction which determines the value of $I$ increases with decreasing bandwidth, but the relevant bandwidth is not W, but the width of the 2*p*(O) band from which the defect band is derived [51]. Thus, a value of $I \sim 1$ eV is plausible.

A last point to emphasise here is that the moment is not limited by the magnitude of the charge transfer. It reflects the occupancy of the defect band, which becomes split as electrons are transferred to or from the charge reservoir. The moment may well exceed the number of transferred electrons. The charge reservoir just serves to tip the energy balance of the defect band in favour of Stoner splitting.

Finally, we consider the magnetization process of the sparse Stoner ferromagnet. Magnetocrystalline anisotropy is expected to be negligibly small in the materials we are considering, for two reasons: i) the unpaired spins are mostly associated with the ligand, oxygen or boron, where the spin-orbit coupling is small, or with electrons trapped at oxygen vacancies (F-centres) or grain boundaries where it is absent and ii) if there is a local direction for the crystal field determined by the local symmetry created by the defects, this direction will vary from site to site, and the anisotropy averages to zero provided it is small to begin with [52]. The resulting coercivity is negligibly small.

The micromagnetic state in these materials will therefore be governed by magnetic dipole interactions, which depend on the shape of the percolating magnetic regions where the unpaired spins are to be found. The ground state will not be strictly ferromagnetic because the local ferromagnetic axis wanders over a macroscopic sample in response to the competition between the local dipole field and the exchange stiffness. There is no tendency to form collinear ferromagnetic domains separated by domain walls when the anisotropy is negligible. For example, if the ferromagnetic regions with the spin-polarized electrons are the grain boundaries as suggested in Fig 4d. the magnetization will tend to lie in the plane of the grain boundary, provided the grain size is greater than the exchange length. The exchange length for Ni for example is 5.1 nm [53]. The magnetization process involves straightening out the ferromagnetic order, which means overcoming the local dipole fields. This process will scale with the saturation magnetization, which is practically independent of temperature well below the Curie temperature. Any residual coercivity due to the dipole interactions will also be temperature-independent, as seen in Fig 2e). A simple calculation for macrospins subject to a local dipole field $M_s$ with random direction and gives the fit to the magnetization curve shown in Fig. 2e. The approach to saturation is determined by the value of the magnetization $M_s$ of the ferromagnetic regions. It may be fitted to within 2-3 percent by a simple tanh function

$$M/M_0 = \tanh(H/H_0) \qquad (6)$$

with $H_0 = 0.16 M_s$. The average value of $H_0$ for our $TiO_2$ samples is 125 kA m$^{-1}$, and the average magnetization of the thin films $M_0$ is 10 kA m$^{-1}$ [Fig.2f]. It follows that the average ferromagnetic volume is just 1.3%. An extensive data mining exercise on more than 200 samples of dilute magnetic oxide thin films and ceramic nanoparticles shows that only 1 – 5% of the volume of the films, and 1 – 10 ppm of the volume of the particles is magnetically ordered.

## 5. Conclusions

Based on our study of the magnetism of Fe-doped $TiO_2$, and recent results by others on related materials [29,44,48], we have proposed a new theory for the origin of ferromagnetic-like behaviour of thin films of dilute magnetic oxides and other, $d^0$ materials where there is no trivial explanation in terms of clusters of a ferromagnetic impurity phase. Unlike our earlier impurity band model with ordered 3$d$ local moments and effective Heisenberg exchange within magnetic polarons [21], the 3$d$ dopant cations do not contribute directly to the magnetism in the present model. Here, Stoner-type ferromagnetism is associated with electrons in percolating defect structures such as grain boundaries, where they form a spin polarized quasi 2-dimensional electron gas, which eludes detection by techniques such as small-angle neutron scattering or XMCD on the dopant or oxygen sites [43,44], The ferromagnetic volume, where the defects are located is remarkable small, ≈ 1% of the film volume. Charge transfer to or from a charge reservoir into a narrow, defect-related band can give rise to the inhomogeneous Stoner-type wandering axis ferromagnetism which reproduces the unusual magnetic properties of these systems — high Curie temperature, anhysteretic temperature-independent magnetization curves, a metallic or insulating ferromagnetic-like ground state and a moment that may exceed that of the dopant cations. All the useful magnetic effects related to spin-orbit coupling are expected to be weak, but the spin polarization of metallic and half-metallic charge transfer ferromagnets may provide some opportunities for applications, particularly if the ferromagnetic axis can be discouraged from wandering by ordering the defects in such a way as to eliminate the random character of the dipole fields. The challenge is now to control the defects so as to deliver the sparse high-temperature ferromagnetism to where it may be useful.

**Acknowledgements**

This work was supported by Science Foundation Ireland as part of the CINSE and MANSE programs. K. Paul was supported by the SFI EUREKA scheme. We are grateful to David Edwards for helpful discussions.

**References**

[1] Matsumoto Y, Murakami M, Shono T, Hasegawa T, Fukumura T, Kawasaki M, Ahmet P, Chikyow T, Koshihara S Y and Koinuma H 2001 *Science* **291,** 854-856

[2] Coey, J M D 2006 *Curr. Opin. Solid State and Mater. Sci.* **10** 83

[3] Chambers S A 2006 *Surf. Sci. Rep.* **61** 345

[4] Garitaonandia J S, Insausti M, Goikolea E, Suzuki M, Cashion J D, Kawamura N, Ohsawa H, De Muro I G, Suzuki K, Plazaola F and Rojo T 2008 *Nano Lett.* **8,** 661

[5] Liou Y, Su P W and Shen Y L 2007 *Appl. Phys. Lett.* **90** 182508

[6] Esquinazi P, Spemann D, Höhne R, Setzer A, Han K –H and Butz T 2003 *Phys. Rev. Lett.* **91** 227201

[7] Ueda K, Tabata H, and Kawai T 2001 *Appl. Phys. Lett.* **79** 988

[8] Coey J M D, Douvalis A P, Fitzgerald C B, and Venkatesan M 2004 *Appl. Phys. Lett*. **84** 1332.

[9] Kale S N, Ogale S B, Shinde S R, Sahasrabuddhe M, Kulkarni V N, Greens R L and Venkatesan T 2003  *Appl. Phys. Lett*. **82**, 2100

[10] Philip J, Punnoose A, Kim B I, Reddy K M, Layne S, Holmes J O, Satpati B, Leclair P R, Santos T S and Moodera J S 2006 *Nature Mater.* **5** 298

[11] Wang Z, Wang W, Tang J, Tung L D, Spinu L and  Zhou W 2003 *Appl. Phys. Lett*. **83**, 518

[12] Hong N H 2004 *Phys. Rev. B*  **70**, 195204.

[13] Kim Y J, Thevuthasan S, Droubay T, Lea A S, Wang C M, Shutthanandan V, Chambers S A, Sears R P, Taylor B and Sinkovic B  2004 *Appl. Phys. Lett*. **84** 3531

[14] Suryanarayanan R, Naik V M, Kharel P, Talagala P and Naik R 2005 *J. Phys. Condens. Matter*. **17** 755

[15] Orlov A F, Perov N S, Balagurov L A, Konstantinova A S and Yarkin D G 2007 *JETP Lett.* **86** 352

[16] Chan K Y S and Goh G K L 2008 *Thin Solid Films* **512** 5582

[17] Rykov A I, Nomura K, Sakuma J, Barrero C, Yoda Y and Mitsui T 2008 *Phys. Rev B.* **77** 014302

[18] Mudarra Navarro A M, Bilovola V, Cabrera A F, Rodríguez Torresa C E and Sánchez F H 2009 *Physica B: Condens. Matter.* **404** 2838

[19] Griffin K A, Pakhomov A B, Wang C M, Heald S M, Krishnan K M 2005 *Phys. Rev. Lett*. **94**, 157204

[20] Zhao T, Shinde S R, Ogale S B, Zheng H, Venkatesan T, Ramesh R and Sarma S D 2005 *Phys. Rev. Lett*. **94** 126601

[21] Coey J M D, Venkatesan M and Fitzgerald C B 2005 *Nature Mater.* **4** 173

[22] Ogale S B, Choudhary R J, Buban J P, Lofland S E, Shinde S R, Kale S N, Kulkarni V N, Higgins J, Lanci C, Simpson J R, Browning N D, Das Sarma S, Drew H D, Greene R L and Venkatesan T 2003 *Phys. Rev. Lett.* **91** 77205

[23] Coey J M D 2005 *J. Appl. Phys*. **97** 10D313

[24] Venkatesan M, Fitzgerald C B and Coey J M D 2004 *Nature* **430** 630

[25] Dorneles L S, Venkatesan M, Moliner M, Lunney J G and Coey J M D 2004 *Appl. Phys. Lett*., **85**, 6377.

[26] Hong N H, Sakai J, Poirot N and Brize V 2006 *Phys. Rev. B*. **73** 132404

[27] Yoon S D, Chen Y, Yang A, Goodrich T L, Zuo X, Arena D A, Ziemer K, Vittoria C and Harris V G 2006 *J. Phys.: Condens. Matter*. **18**, L355

[28] Sundaresan A, Bhargavi R, Rangarajan N, Siddesh U and Rao C N R 2006 *Phys. Rev. B* **74** 161306

[29] Zhou Z, Čižmár E, Potzger K, Krause M, Talut G, Helm M, Fassbender J, Zvyagin S A, Wosnitza J and Schmidt H 2009 *Phys. Rev. B*. **79** 113201

[30] Lawes G, Risbud A S, Ramirez A P and Seshadri R. 2005 *Phys. Rev. B*. **71** 045201

[31] Rao C N R and Deepak F L 2005 *J. Mat. Chem*. **15** 573

[32] Kaspar T C, Heald S M, Wang C M, Bryan J D, Droubay T, Shutthanandan V, Thevuthasan S, McCready D E, Kellock A J, Gamelin D R and Chambers S A 2005 *Phys. Rev. Lett*. **95** 217203

[33] Chambers S A, Droubay T, Wang C M, Lea A S, Farrow R F C, Folks L, Deline V and Anders S 2003 *Appl. Phys. Lett*. **82** 1257

[34] Abraham D W, Frank M M and Guha S 2005 *Appl. Phys. Lett*. **87** 252502

[35] García M A, Fernández Pinel E, De la Venta J, Quesada A, Bouzas V, Fernández J F, Romero J J, Martín-González M S, Costa-Kramer, J L 2009 *J. Appl. Phys*. **105** 013925

[36] Stamenov P and Coey J M D 2006 *Rev. Sci. Instruments* **77** 015106

[37] Coey J M D, Wongsaprom K, Alaria J and Venkatesan M. 2009 *J. Phys.D: Appl. Phys*. **41** 134012

[38] Pascual J, Camassel J and Mathieu H 1978 *Phys. Rev. B*. **18** 5606

[39] Liljequist D 1998 *Nucl. Inst. Meth*. **142** 295

[40] Cabrera A F, Rodríguez Torres C E, Errico L and Sánchez F H 2006 *Physica* B **384** 345.

[41] Xiaoyan P, Dongmei J, Yan L and Xueming Ma 2006 *J. Magn. Mag. Mater.* **305** 388.

[42] Barla A, Schmerber G, Beaurepaire E, Dinia A, Bieber H, Colis S, Scheurer F, Kappler J -P, Imperia P, Nolting F, Wilhelm F, Rogalev A, Müller D and Grob J J 2007 *Phys. Rev. B.* **76** 125201.[43] Tietze T, Gacic M, Schütz G, Jakob G, Brück S and Goering E 2008, *New J. Phys.* **10** 055009

[44] Mudarra Navarro A M, Bilovol V, Cabrera A F, Rodríguez Torres C E and Sánchez F H. 2009 *Physica B* **404** 2838.

[45] Gallego S, Beltrán J I, Cerdá J and Muñoz M C 2005 *J. Phys.: Condens. Matter.* **17** L451.

[46] Vager Z and Naaman R 2004 *Phys. Rev. Lett*. **92** 087205

[47] Hernando A, Crespo P, Garcia M A 2006 *Phys. Rev. Lett*. **96** 057206

[48] Straumal B B, Mazilkin A A, Protasova, S G, Myatiev A A, Straumal P B, Schütz G, van Aken P A, Goering E and Baretzky B. 2009 *Phys. Rev. B.* **79** 205206

[49] Janak J F 1977 *Phys. Rev. B*. 16 255

[50] Behan A J, Mokhtari A, Blythe H J, Score D, Xu X-H, Neal J R, Fox A M and Gehring G A 2008 *Phys. Rev. Lett.* **100** 047206

[51] Edwards D M and Katsnelson M I 2006 *J. Phys.C:Cond. Matter*. **18** 7209

[52] Chi M C and Alben R 1977 *J. Appl. Phys*. **48** 2987

[53] J. M. D. Coey 2010 *Magnetism and Magnetic Materials*, Cambridge University Press, 638 pp.

Table 1 Fit Parameters for Mössbauer spectra. Isomer shifts are relative to an α-Fe absorber.

| Fe (at%) | | Isomer shift (mm s$^{-1}$) | Quadrupole Splitting (mm s$^{-1}$) | $B_{hf}$ (T) | Area (%) |
|---|---|---|---|---|---|
| 1 | $Fe^{2+}$ | 0.96 | 2.11 | 0.0 | 78 |
|   | $Fe^{3+}$ | 0.40 | 0.56 | 0.0 | 22 |
| 3 | $Fe^{2+}$ | 0.96 | 2.11 | 0.0 | 13 |
|   | $Fe^{3+}$ | 0.41 | 0.60 | 0.0 | 87 |
| 5 | $Fe^{2+}$ | 0.42 | -0.20 | 49.0 | 60 |
|   | $Fe^{3+}$ | 0.39 | 0.53 | 0.0 | 40 |

**Figure captions**

Figure 1 X-ray diffraction pattern of a) a blank R-cut sapphire substrate. b) a 1% Fe-doped oxidized TiO$_2$ film grown in oxygen pressure of 10$^2$ Pa.

Figure 2 Magnetic properties of iron doped rutile films: a) Room temperature magnetization curves of an undoped TiO$_2$ film and oxidized 1% and 3% Fe doped TiO$_2$ films on sapphire b) The same data after subtracting the diamagnetic contribution of the substrate c) Magnetic moments of oxidized films made with normal iron or $^{57}$Fe plotted as Bohr magnetons per Fe d) $\chi_m$ vs 1/T plots for the oxidized, 1% Fe-doped film e) Magnetization curves for 1% Fe-doped TiO$_2$ film at 4 K and 300 K and f) scatter plot of magnetization vs $H_0$ (see text) for the magnetic iron-doped rutile films shown in Fig.2c.

Figure 3 Conversion electron Mössbauer spectra at 300K of oxidized TiO$_2$ films doped with a) 1%, b) 3%, c) 5% of iron ($^{57}$Fe) and d) Mössbauer spectra of metallic iron

Figure 4 Possible distributions of defects: a) Random distribution of point defects at the percolation threshold b) spinodal decomposition c) interface defects d) grain boundary defects.

Fig. 5 Schematic of the idea underlying charge transfer ferromagnetism. Electron transfer to or from the charge reservoir to the defect band leads to the fulfilment of the Stoner criterion, and spontaneous ferromagnetic splitting occurs [45].

Fig. 6 The phase diagram for the charge transfer ferromagnetism model calculated for fixed $I/W = 1$ and $n_0/W = 0.4$. The electronic structure of the different regions is indicated

in the caption. The Ferromagnetic regions are coloured. Metallic and half-metallic regions are shaded. The red line marks half filling of the impurity band. The occupancy of the spin split impurity band for a cut at $U/W = 0.4$ is illustrated by the series of diagrams at the top, where the zero of energy is set at $E_F$. The values of $N_{tot}/W$ are indicated (red is spin up, blue is spin down).

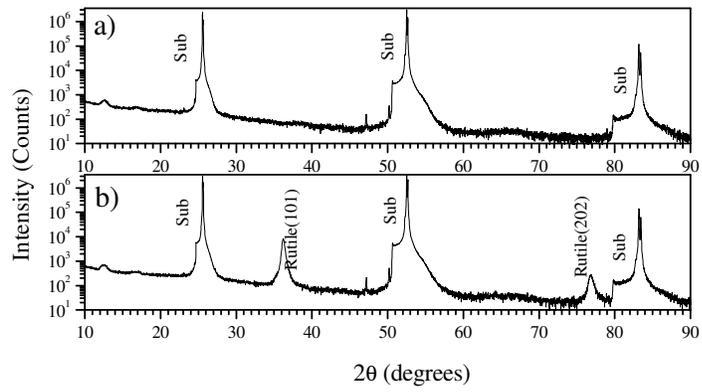

Fig.1

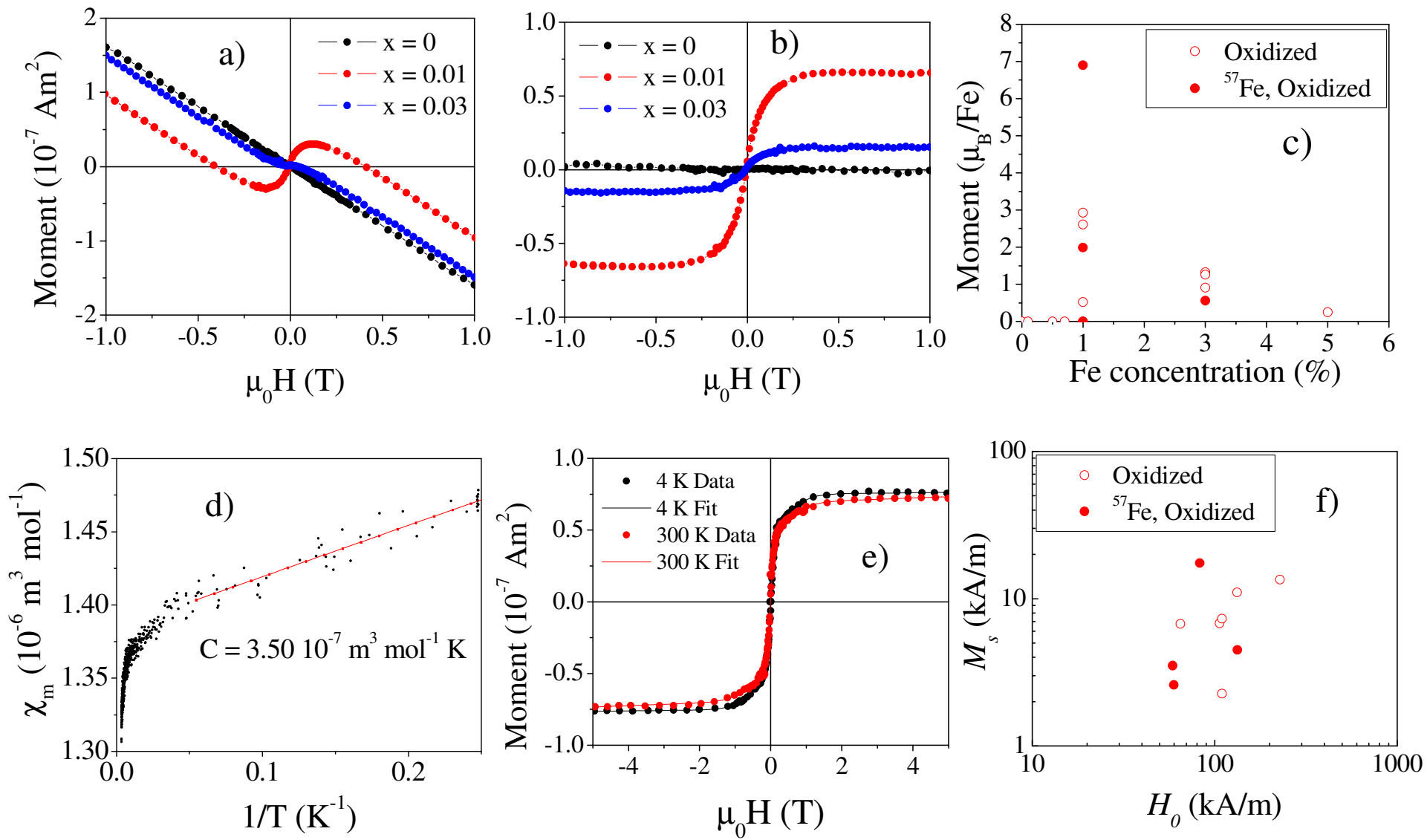

Fig.2

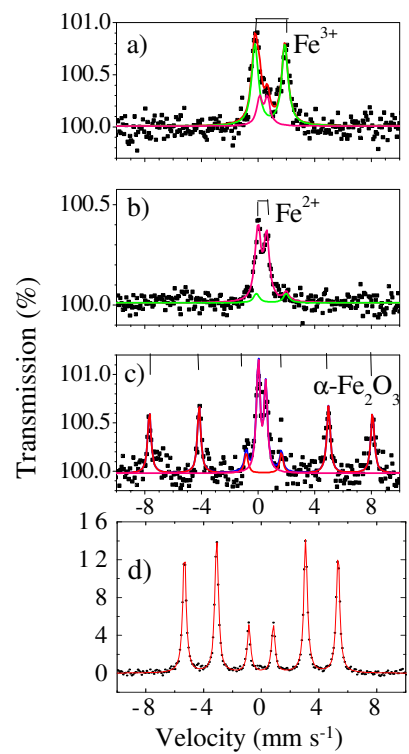

Fig.3

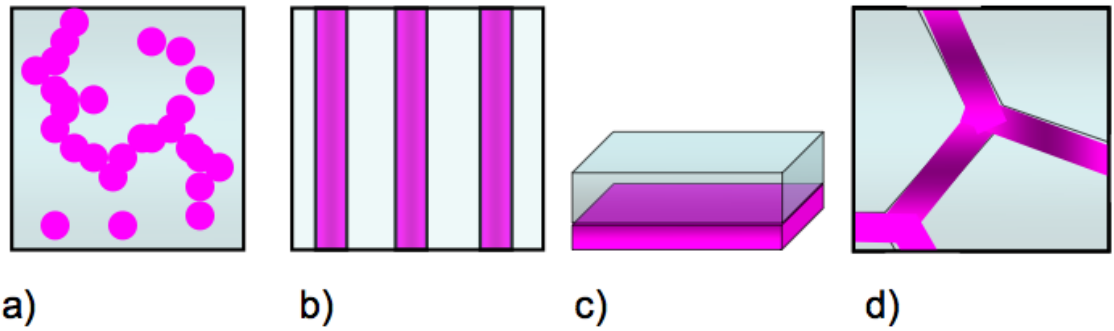

Fig.4

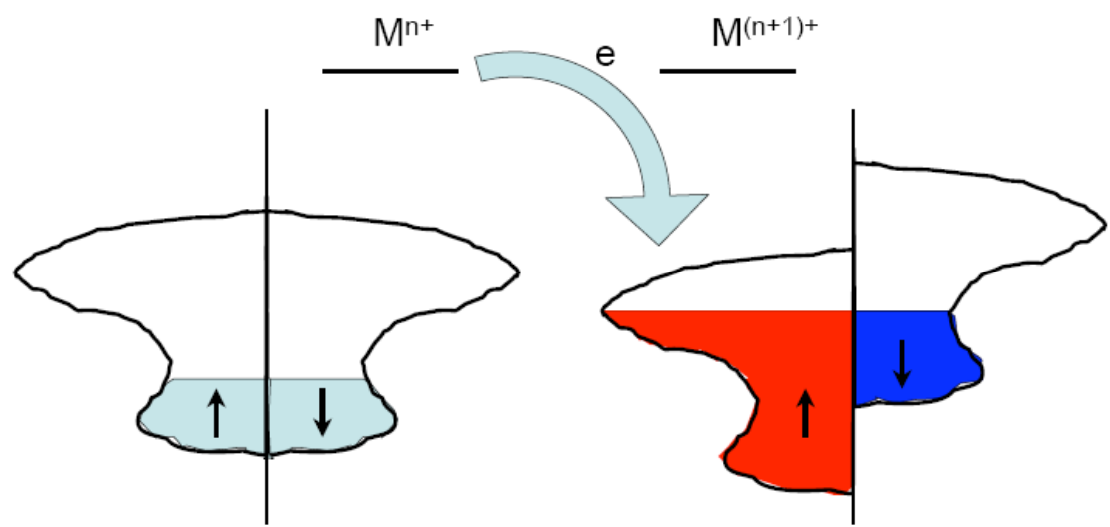

Fig.5

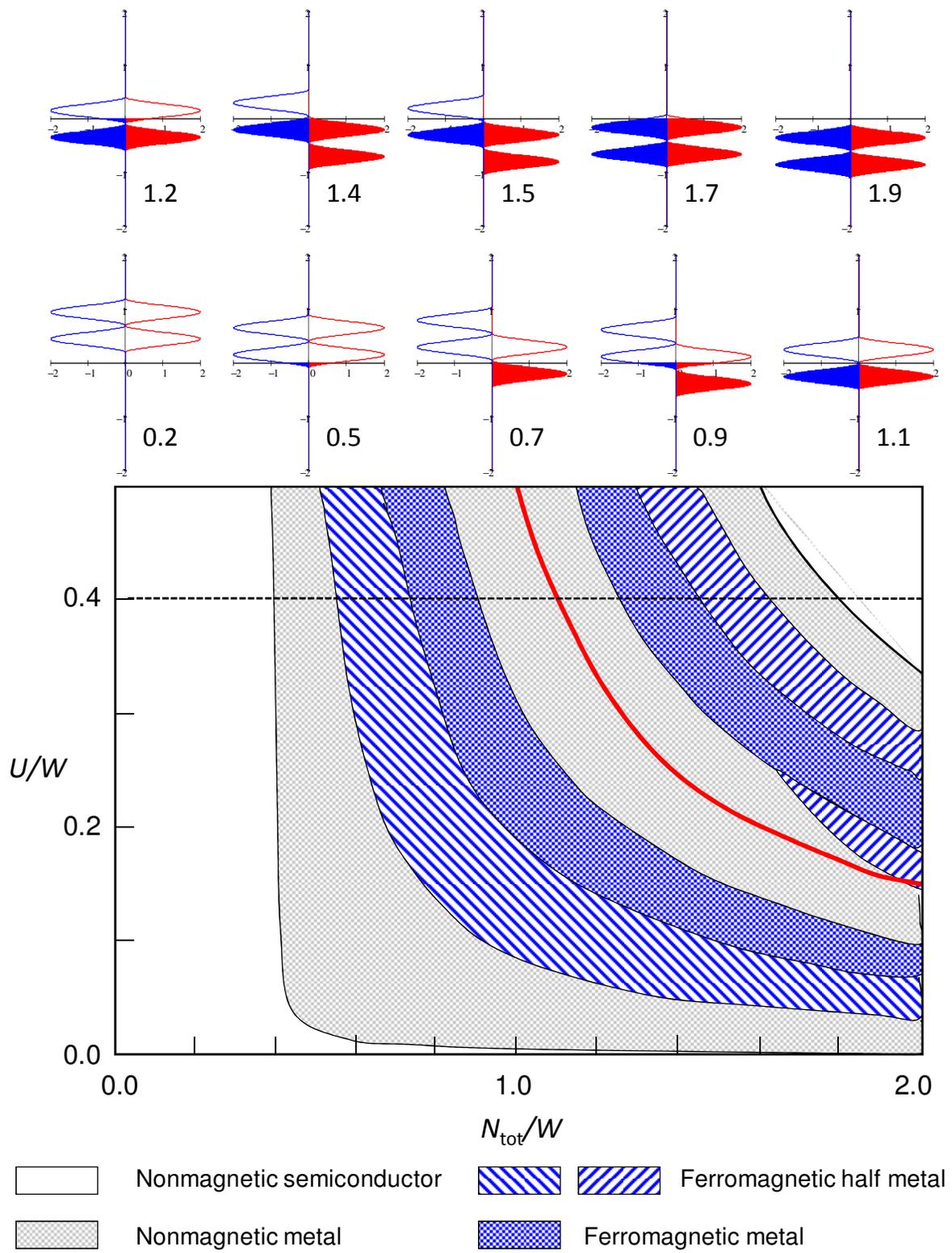

Fig.6